\documentclass{optica-article}

\journal{opticajournal} % for journals or Optica Open

\articletype{Research Article}

\usepackage{listings}
\usepackage{xcolor}
\usepackage{siunitx}
\usepackage{lineno}
\usepackage{bm}
\begin{document}
% MATLAB code style for listings package
\lstdefinestyle{MATLABStyle}{
    language=Matlab,
    basicstyle=\ttfamily\footnotesize,
    keywordstyle=\color{blue},
    commentstyle=\color{green},
    stringstyle=\color{red},
    numbers=left,
    numberstyle=\tiny\color{gray},
    stepnumber=1,
    numbersep=10pt,
    backgroundcolor=\color{lightgray!10},
    showspaces=false,
    showstringspaces=false,
    breaklines=true,
    breakatwhitespace=true,
    tabsize=4
}

\title{Independent amplitude and phase modulation using two sequential regions of one phase-only SLM}

\author{Marius Constantin Chirita Mihaila,\authormark{1,*}, Mikuláš Fiala,\authormark{1} and Martin Kozák,\authormark{1,*} }

\address{\authormark{1}Charles University, Faculty of Mathematics and Physics, Ke Karlovu 3, 121 16 Prague 2\\}

\email{\authormark{*}marius.chirita@matfyz.cuni.cz} %% email address is required; see note below about the corresponding author designation
\email{\authormark{*}m.kozak@matfyz.cuni.cz}
% use {asbstract*} to suppress the copyright line. Copyright information will be added in production

\begin{abstract*} 
Liquid-crystal spatial light modulators (SLMs) are widely used for programmable wavefront control but are typically operated as phase-only devices, limiting applications that require independent amplitude and phase shaping. Here we demonstrate full complex-field modulation using a single phase-only SLM by implementing two sequential modulation planes on different regions of the same device. The phase retardance introduced by the first SLM region is converted into amplitude modulation by a polarizer placed in the beam path, while the second region compensates the associated phase offset and imposes the required phase distribution. The field from the first region is imaged onto the second, enabling complex-field synthesis without a second modulator. We validate the approach by generating Bessel--Gaussian beams, helical-phase fields, and arbitrary focal-plane intensity patterns. This single-SLM platform provides a compact route to programmable complex wavefront engineering for structured illumination, holography, and electron--light interaction experiments.

\end{abstract*}

%%%%%%%%%%%%%%%%%%%%%%%%%%  body  %%%%%%%%%%%%%%%%%%%%%%%%%%
\section{\label{sec:level1}Introduction}

Spatial light modulators (SLMs) have become central tools for programmable wavefront engineering because they allow optical fields to be modified in a computer-controlled and reconfigurable manner~\cite{forbes2016creation}. In adaptive optics, phase-only liquid-crystal SLMs can compensate wavefront distortions and generate controlled aberration modes, including Zernike modes, by imposing an appropriate spatial phase retardance on the incident field~\cite{love1997wave,lakshminarayanan2011zernike}. More broadly, SLMs provide a flexible platform for optical microscopy and structured illumination, where they can shape the illumination field or act as programmable Fourier-plane filters in the imaging path~\cite{maurer2011spatial}. In these applications, phase-only modulation is often sufficient: the SLM changes the optical path length locally, thereby controlling propagation, focusing, aberrations, and interference without directly modifying the incident amplitude.

The same principle underlies phase-only computer-generated holography (CGH). Iterative phase-retrieval algorithms, including the Gerchberg--Saxton algorithm~\cite{gerchberg1972practical} and related methods analyzed by Fienup~\cite{fienup1982phase}, make it possible to calculate phase masks whose propagated intensity approximates a desired target pattern, as reviewed recently in Ref.~\cite{dong2023phase}.
 Such approaches have enabled a wide range of focal-plane shaping experiments with phase-only devices. For example, phase-only holography combined with spatially chirped ultrafast pulses has been used to generate arrays of spatiotemporal foci and extended excitation patterns with simultaneous spatial and temporal control~\cite{sun2018four}. These demonstrations illustrate the strength of phase-only SLMs: even without direct amplitude modulation, propagation and interference can redistribute optical power into highly nontrivial focal structures.

However, phase-only holography has an intrinsic limitation. When only the target intensity is constrained, the phase of the reconstructed field is generally not prescribed but instead remains an algorithmic degree of freedom~\cite{gerchberg1972practical,fienup1982phase}. For coherent fields, this uncontrolled phase can lead to speckle, nonuniformity, and reconstruction artifacts. Several methods have therefore been developed to improve phase-only holographic projection. Examples include high-accuracy hologram-generation methods using optimized quadratic initial phase distributions~\cite{pang2016high}, double-constraint Gerchberg--Saxton algorithms with Fourier-plane amplitude and phase constraints~\cite{chang2015speckle}, and random-phase-free approaches for reducing speckle noise~\cite{shimobaba2015improvement}. Recent developments have extended this line of work through more explicit control of the target-plane phase. These include bandwidth-constrained phase-hologram optimization with quadratic phase
initialization to suppress speckle and reconstruction artifacts
~\cite{chen2021bandwidth}, as well as vortex-annihilation strategies
that reduce phase singularities and produce smoother target-phase
distributions for speckle-suppressed reconstruction
~\cite{sui2025vortex}. These methods improve the quality of intensity reconstructions, but they do not in general provide direct and independent prescription of both the amplitude and the phase of the optical field in the target plane.

Independent amplitude and phase control is especially useful for structured beams whose defining properties are encoded in the complete optical field. Bessel beams are a representative example. Ideal scalar Bessel beams are nondiffracting solutions of the wave equation, and experimentally realizable finite-aperture Bessel or Bessel--Gaussian beams retain extended depth of focus and characteristic radial ring structures~\cite{durnin1987exact,mcgloin2005bessel}. When such beams are combined with an azimuthal phase, the field acquires a vortex-like phase singularity, so that the desired optical state is specified not only by its radial intensity profile but also by its phase topology. Analogous ideas have been extended to electron optics, where electron vortex beams and electron Bessel beams have been generated using holographic electron-optical elements~\cite{verbeeck2010production,grillo2014}. 

A further motivation for optical complex-field synthesis comes from light-based electron-beam shaping. Tailored optical fields can imprint transverse phase profiles onto free-electron wave functions in free space, enabling concepts such as aberration correction and programmable electron focal profiles~\cite{garcia2021optical,uesugi2021electron,uesugi2022properties,chirita2022transverse,chirita2025design,chirita2025light,mihaila2025light,nekula2025compensating}. More generally, coherent electron--light interactions in free space, optical near fields, and nanophotonic structures are now recognized as a route to attosecond electron-wavefunction control, quantum nanophotonics with free electrons, and stimulated low-energy electron--light interactions~\cite{vanacore2018attosecond,rivera2020light,streshkova2024monochromatization,chlouba2023coherent,garcia2025roadmap,chahshouri2026stimulated}. These applications require optical fields with accurately engineered transverse structure at the electron--light interaction plane, making compact and flexible optical amplitude-and-phase shaping a useful enabling technology.

Amplitude-only modulation can attenuate the field
locally, but it cannot directly impose an independent phase profile;
phase-only modulation can efficiently redirect light, but it cannot
directly prescribe an arbitrary local amplitude. Amplitude holography
can nevertheless encode arbitrary complex fields in a selected
diffraction order, although it is typically accompanied by strong
zero and conjugate diffraction orders and reduced optical efficiency.
Direct local complex-field synthesis therefore requires independent
control of both amplitude and phase. Several approaches have addressed this problem. Two cascaded phase-only SLMs can provide arbitrary and independent control of spatial amplitude and phase~\cite{zhu2014arbitrary}. Alternatively, complex fields can be encoded into a
single phase-only hologram by controlling the local diffraction
efficiency and correcting the associated phase. In conventional
implementations, the desired field is isolated in a selected diffraction
order~\cite{arrizon2007pixelated,bolduc2013exact}, while on-axis implementations have also
been demonstrated in which the complex-amplitude-modulated field
propagates coaxially with the zeroth order~\cite{rodriguez2024onaxis}.

Single-SLM dual-phase schemes have demonstrated complex-field generation using two independently addressed regions of one modulator~\cite{baliyan2023generation}. However, the ideal double-phase model relies on accurate one-to-one coordinate mapping between the two phase masks, which generally requires unit-magnification imaging between the corresponding SLM regions~\cite{fienup2024fourier}. Deviations from this condition can reduce interference fidelity and introduce residual amplitude noise or speckle in the reconstructed field. 

 Other holographic methods have achieved near-perfect reconstruction and amplitude-and-phase control for optical tweezers~\cite{jesacher2008near,jesacher2008full}, but focusing the beam onto the SLM may impose damage-threshold constraints in high-power applications. Related double-pass SLM geometries have also been used for arbitrary polarization control~\cite{clegg2013double}. Independent amplitude and phase control with a single phase-only SLM has also been demonstrated using a high-spatial-frequency phase grating and Fourier-plane spatial filtering~\cite{bagnoud2004independent}. In that approach, the difference between two alternating phase values controls the transmitted amplitude, while their mean determines the phase of the filtered zeroth-order field. The method requires only one SLM interaction, but relies on a carrier grating, precise pinhole alignment, and low-pass spatial filtering. In comparison, the two-region scheme introduced below avoids carrier gratings and diffraction-order filtering, but requires accurate registration, additional polarization optics, and relay-phase compensation.

Here we demonstrate independent amplitude and phase shaping using two
spatially separated regions of a single phase-only SLM. The first
region, together with a polarizer, converts the programmed phase
retardance into amplitude modulation. The field is then imaged with unit
magnification onto the second region, which compensates the phase
introduced during amplitude encoding and imposes the desired output
phase. This arrangement preserves a direct coordinate mapping between
the amplitude and phase masks while replacing two separate modulators
with a single device.

The central advances of this work are twofold. First,
the single-device architecture provides compact, directly mapped
complex-field control without requiring tight focusing onto the SLM
surface, making it potentially advantageous for high-peak-power
operation. Second, we combine this architecture with an
inverse-propagation workflow to generate prescribed optical fields
directly in the focal region~\cite{voelz2011computational}. We validate
independent amplitude and phase control using Bessel--Gaussian beams
with and without an added helical phase, and further demonstrate
arbitrary focal-plane intensity landscapes with close agreement between
experiment and simulation. In particular, we experimentally generate
helical-ramp and Zernike-like focal-intensity distributions that can
serve as programmable ponderomotive phase elements for free-electron
beams. The former offers a route to electron-vortex generation, while
the latter enables tailored correction of electron-optical aberrations,
analogous to programmable aberration correction in light optics~\cite{love1997wave}. The method therefore extends beyond generic complex-field synthesis and provides a practical platform for
structured illumination~\cite{hohenester2025optimizing},
holography~\cite{forbes2016creation,bolduc2013exact}, and application-specific optical shaping of free-electron wavefunctions~\cite{ferrari2025realization,garcia2025roadmap}.

\section{\label{sec:level2}Experimental concept and theoretical framework}

The concept of the experiment is to use two spatially separated regions of a single phase-only SLM as two sequential modulation planes. Although each region can only impose a phase retardance, the first region, together with a polarizer placed in the beam path, provides amplitude modulation, while the second region provides phase modulation, enabling control of the full complex optical field. In this way, the first SLM region is used to encode the desired field amplitude, while the second SLM region independently imposes the remaining phase required to synthesize a prescribed complex field in the pupil plane.

\begin{figure}[htp]
   \centering      
   \includegraphics[width=7.0cm]{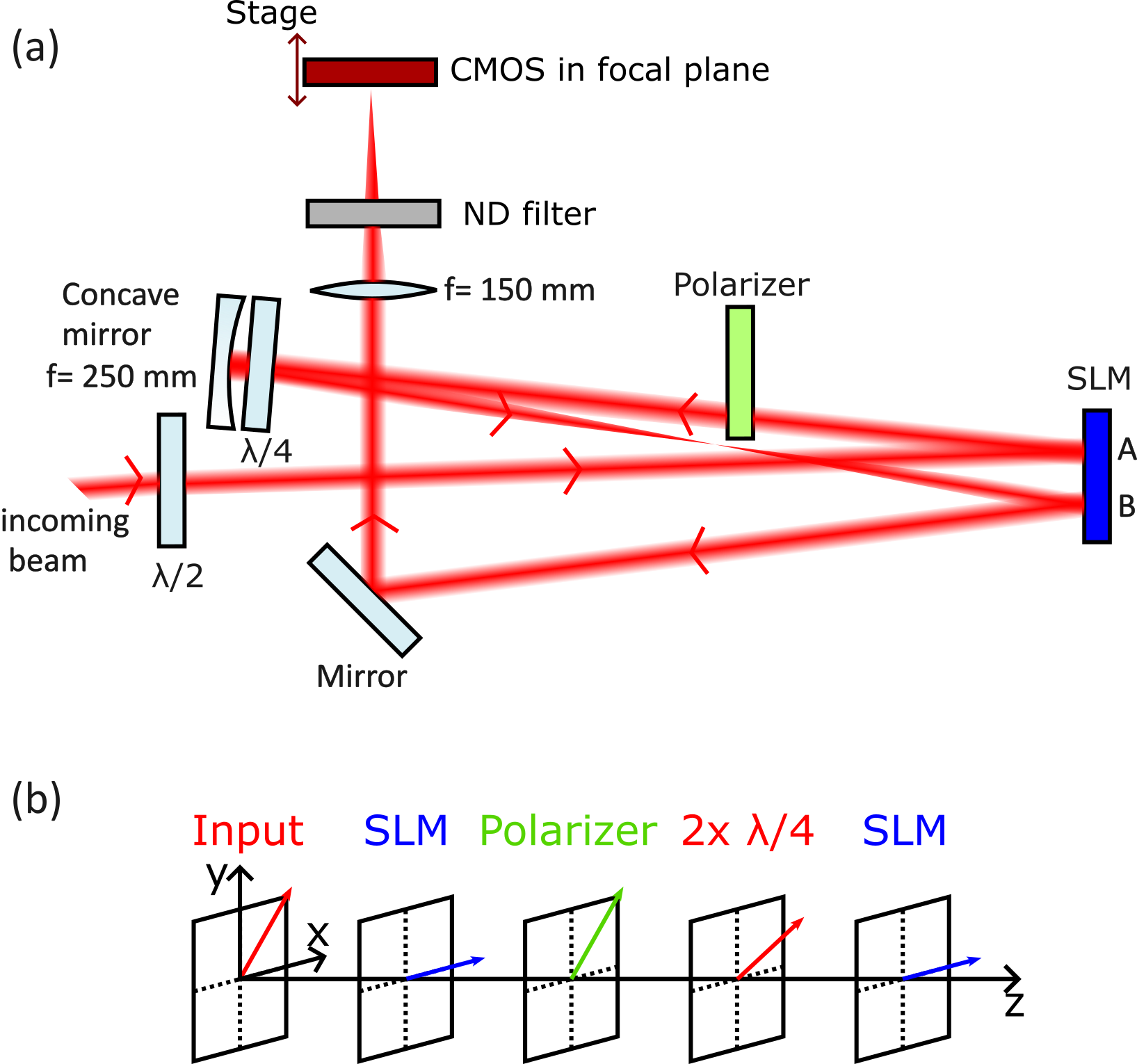}
   
   \caption{
Experimental implementation for independent amplitude and phase control using a single SLM.
(a) The input beam is polarized at $45^\circ$ with respect to the active axis of the SLM and illuminates region A. After projection by a linear polarizer, the SLM-induced retardance is converted into amplitude modulation. Region A is imaged with unit magnification onto region B by a folded single-mirror relay based on a concave mirror with $f_m=\SI{250}{mm}$. A double pass through the quarter-wave plate rotates the polarization onto the active axis of the SLM, allowing region B to impose the desired phase modulation. A compensating quadratic phase is superimposed on the phase mask displayed on region B to cancel the wavefront curvature introduced by the relay and produce a collimated output beam. The subsequent physical lens with $f_F=\SI{150}{mm}$ performs the Fourier transformation, and the CMOS camera is positioned at its back focal plane.
(b) Polarization evolution along the optical path. The input field is polarized at $45^\circ$, such that only its horizontal component is phase-modulated by the SLM, while the orthogonal component remains ideally unchanged. The panel also shows the polarization projection by the linear polarizer and the subsequent rotation after the double pass through the quarter-wave plate, after which the beam is horizontally polarized along the working axis of the SLM before incidence on region B.
}

   \label{fig:experimental setup}
\end{figure}

The experimental implementation is shown in Fig.~\ref{fig:experimental setup}. The incident beam is linearly polarized at \(45^\circ\) with respect to the active axis of the SLM and illuminates region A. The phase retardance applied in this region creates a spatially dependent phase difference between the two polarization components. The phase retardance applied in region A produces a spatially varying polarization ellipticity; a linear polarizer aligned with the incident polarization direction then converts this ellipticity into an effective amplitude modulation by transmitting only the corresponding polarization component. The modulated field is then imaged with unit magnification onto region B of the same SLM, so that the transverse coordinates in both regions can be identified with the same pupil-plane coordinates \((x_p,y_p)\). A double pass through the quarter-wave plate rotates the polarization onto the active axis of the SLM, allowing region B to apply a second independently programmable phase pattern. The resulting complex field is then focused by a lens onto the camera plane.

\subsection{Complex-field synthesis using two phase-only SLM regions}

The role of the setup in Fig.~\ref{fig:experimental setup} can be expressed directly in terms of the two phase patterns applied to regions A and B. After the first interaction with SLM region A and transmission through the linear polarizer, the modulated and unmodulated polarization components are combined into the transmitted polarization state. This gives an amplitude factor proportional to
$\cos[\phi_{\mathrm{A}}(x_p,y_p)/2]$, together with an additional phase contribution
$\phi_{\mathrm{A}}(x_p,y_p)/2$.
After compensation of the quadratic phase introduced by the imaging relay, region B contributes the effective phase
$\phi_{\mathrm{B}}(x_p,y_p)$.
Thus, after propagation through the two independently addressed regions, the scalar complex optical field can be written as~\cite{zhu2014arbitrary}
\begin{equation}
  E_{\mathrm{out}}(x_p,y_p)
  =
  A_0
  \cos\!\left[\frac{\phi_{\mathrm{A}}(x_p,y_p)}{2}\right]
  \exp\!\left\{
  i\left[
  \frac{\phi_{\mathrm{A}}(x_p,y_p)}{2}
  +
  \phi_{\mathrm{B}}(x_p,y_p)
  \right]
  \right\},
  \label{eq:two_slm_field}
\end{equation}
where $A_0$ is the real-valued amplitude of the incident field,
$\phi_{\mathrm{A}}(x_p,y_p)$ is the phase distribution encoded on region A, and
$\phi_{\mathrm{B}}(x_p,y_p)$ is the effective phase contribution of region B after compensation of the quadratic phase introduced by the imaging relay.
Equation~\eqref{eq:two_slm_field} shows that region A controls the output amplitude through the factor
$\cos[\phi_{\mathrm{A}}(x_p,y_p)/2]$, while the total output phase is determined by the sum of the phase generated during amplitude encoding and the phase imposed by region B. A Jones-matrix derivation of Eq.~\eqref{eq:two_slm_field} is given in Appendix~\ref{sec:A}.

The desired complex field in the SLM/pupil plane is written in amplitude--phase form as
\begin{equation}
  E_{\mathrm{pupil}}(x_p,y_p)
  =
  A_0 \mathcal{A}(x_p,y_p)\exp[iP(x_p,y_p)],
  \label{eq:output_field}
\end{equation}
where $\mathcal{A}(x_p,y_p)$ is the amplitude normalized to the incident amplitude $A_0$, and $P(x_p,y_p)$ is the desired phase distribution. Since the scheme redistributes and attenuates the incident field but does not locally amplify it, the normalized amplitude must satisfy
\begin{equation}
  0 \leq \mathcal{A}(x_p,y_p) \leq 1.
  \label{eq:amplitude_constraint}
\end{equation}
Equating Eq.~\eqref{eq:two_slm_field} with Eq.~\eqref{eq:output_field} gives the required phase pattern on region A,
\begin{equation}
  \phi_{\mathrm{A}}(x_p,y_p)
  =
  2\arccos[\mathcal{A}(x_p,y_p)].
  \label{eq:phi1_general}
\end{equation}
Thus, the first SLM region maps the desired normalized amplitude onto an effective amplitude modulation.
The phase contribution introduced by this step is then compensated when choosing the effective phase contribution of region B.
The required second phase mask is
\begin{equation}
  \phi_{\mathrm{B}}(x_p,y_p)
  =
  P(x_p,y_p)
  -
  \arccos[\mathcal{A}(x_p,y_p)].
  \label{eq:phi2_explicit}
\end{equation}

The unit-magnification relay introduces the additional quadratic phase
\begin{equation}
  \phi_{\mathrm{rel}}(x_p,y_p)
  =
  \frac{\pi}{\lambda_L f_m}
  \left(x_p^2+y_p^2\right)
  \label{eq:relay_phase}
\end{equation}
where $f_m=\SI{250}{mm}$ is the focal length of the concave mirror and the transverse coordinates are measured relative to the relay optical axis. The coordinate inversion associated with the unit-magnification imaging is accounted for when constructing the pattern on region B. The relay-induced phase is compensated by adding the opposite quadratic phase to the pattern displayed on region B. The complete phase physically displayed on region B is therefore
\begin{equation}
  \begin{aligned}
    \phi_{\mathrm{B,disp}}(x_p,y_p)
    &=
    \phi_{\mathrm{B}}(x_p,y_p)
    -
    \phi_{\mathrm{rel}}(x_p,y_p)
    .
  \end{aligned}
  \label{eq:phase_B_displayed}
\end{equation}
At the output of region B, the relay-induced quadratic phase and its digitally applied compensation cancel, leaving the effective phase
$\phi_{\mathrm{B}}(x_p,y_p)$. The resulting field is therefore given by Eq.~\eqref{eq:output_field}.

Equations~\eqref{eq:phi1_general} and \eqref{eq:phase_B_displayed} provide the direct prescription for synthesizing an arbitrary complex field in the SLM/pupil plane using two phase-only modulation steps. Region A determines the normalized amplitude
$\mathcal{A}(x_p,y_p)$, while region B compensates both the phase offset introduced by the amplitude-encoding step and the quadratic phase introduced by the imaging relay, while imposing the desired phase $P(x_p,y_p)$. Consequently, the pair of physically displayed phase masks
$\{\phi_{\mathrm{A}},\phi_{\mathrm{B,disp}}\}$ specifies the target field
$A_0\mathcal{A}(x_p,y_p)\exp[iP(x_p,y_p)]$, provided that Eq.~\eqref{eq:amplitude_constraint} is satisfied.

\subsection{Double-phase hologram calculation for arbitrary focal shapes}

The relations above specify how to encode a desired complex field in the pupil plane. To generate a desired field distribution in the focal plane, we first calculate the pupil-plane field that produces it after propagation through the Fourier lens. The target complex field is specified in the focal plane as
\begin{equation}
    E_{\mathrm{tar}}(x_f,y_f)
    =
    A_{\mathrm{tar}}(x_f,y_f)
    \exp[i\Phi_{\mathrm{tar}}(x_f,y_f)].
    \label{eq:focal_target_field}
\end{equation}
The corresponding field in the SLM/pupil plane is obtained by inverse Fourier propagation~\cite{Chen2016ReverseEngineeringFocusShaping} and restriction to the illuminated SLM aperture,
\begin{equation}
    E_{\mathrm{pupil}}(x_p,y_p)
    \propto
    \mathcal{F}^{-1}
    \left\{
    E_{\mathrm{tar}}(x_f,y_f)
    \right\}
    \Pi_R(x_p,y_p),
    \label{eq:focal_to_pupil_field}
\end{equation}
where \(\Pi_R(x_p,y_p)\) is the circular pupil function of radius \(R=\SI{2.8}{mm}\). This value was set by the alignment constraints of the setup, since a larger pupil radius would increase the risk of clipping at the polarizer and during imaging of region A back onto region B of the SLM. The proportionality accounts for the subsequent normalization of the pupil-plane amplitude so that the condition in Eq.~\eqref{eq:amplitude_constraint} is fulfilled.

After normalization, the pupil-plane field in Eq.~\eqref{eq:focal_to_pupil_field} is identified with Eq.~\eqref{eq:output_field}. Its amplitude \(\mathcal{A}(x_p,y_p)\) and phase \(P(x_p,y_p)\) are then inserted into Eqs.~\eqref{eq:phi1_general} and \eqref{eq:phi2_explicit} to obtain the two phase-only masks applied to regions A and B of the SLM. Finally, both masks are wrapped modulo \(2\pi\), converted into additive phase corrections relative to the uniform SLM background phase, and inserted into the two addressed SLM regions. This same back-propagation and double-phase encoding procedure is used for all focal-plane patterns reported below; the different examples differ only in the choice of the target field \(E_{\mathrm{tar}}(x_f,y_f)\). Additional implementation details are provided in Appendix~\ref{app:general_hologram_formalism}.

\section{\label{sec:level3}Results}

We experimentally validate the single-SLM complex-field modulation scheme using the setup shown in Fig.~\ref{fig:experimental setup}. The measurements were performed with a $\lambda_\mathrm{L}=\SI{1030}{nm}$, \SI{280}{fs} laser beam incident on a HOLOEYE PLUTO-2.1 phase-only SLM. At the SLM plane, the input beam had a $1/e^2$ intensity radius of approximately \SI{2}{mm}.

All target fields were generated using the same numerical procedure. First, the desired focal-plane field \(E_{\mathrm{tar}}(x_f,y_f)\) was specified in the focal plane of the Fourier lens. The corresponding pupil-plane field was then obtained by inverse propagation using Eq.~\eqref{eq:focal_to_pupil_field}, restricted to the illuminated aperture, and normalized to satisfy the amplitude constraint in Eq.~\eqref{eq:amplitude_constraint}. The resulting pupil-plane amplitude \(\mathcal{A}(x_p,y_p)\) and phase \(P(x_p,y_p)\) were converted into the two phase masks \(\phi_{\mathrm{A}}(x_p,y_p)\) and \(\phi_{\mathrm{B}}(x_p,y_p)\) using Eqs.~\eqref{eq:phi1_general} and \eqref{eq:phi2_explicit}. Region A of the SLM therefore determines the local field amplitude through the polarization-projection (see Appendix~\ref{Intensity modulation calibration} for details), whereas region B compensates the phase introduced by the amplitude encoding and imposes the desired pupil-plane phase. The spatial resolution of the amplitude-modulation step was characterized separately in the SLM image plane using a reduced input beam diameter, and is therefore reported as a setup-specific diagnostic rather than as the resolution limit of the focal-plane measurements (see Appendix~\ref{sec:spatial_resolution} for details). Numerical simulations were performed by reconstructing the encoded complex pupil field from Eq.~\eqref{eq:two_slm_field}, using the same phase masks as in the experiment, and propagating this field to the focal plane. For the longitudinal intensity maps, the same encoded field was propagated to a sequence of axial planes around the nominal focus using scalar diffraction propagation. No carrier grating or diffraction-order filtering is required. The compensating quadratic phase applied on region B collimates the modulated field, while residual unmodulated light does not acquire this correction, remains divergent at the Fourier lens, and therefore contributes only a weak, broad background in the focal plane.

As a first demonstration, we generated Bessel--Gaussian-type focal fields. The target field was a finite-aperture, super-Gaussian-apodized zeroth-order Bessel field, as described in Appendix~\ref{app:bessel_target}. Figure~\ref{fig:bessel_gauss_helical} compares the simulated and experimentally measured transverse intensity distributions for a Bessel$_0$--Gauss beam and for the same radial field after adding a helical phase. The measured Bessel$_0$--Gauss beam reproduces the expected central maximum and surrounding ring structure. After adding the helical phase, the on-axis phase singularity suppresses the central intensity, producing the annular profile observed in both the simulation and the experiment. This result demonstrates that the setup can impose an azimuthally varying phase independently of the radial amplitude distribution.

\begin{figure}[htp]
   \centering      
   \includegraphics[width=8.0cm]{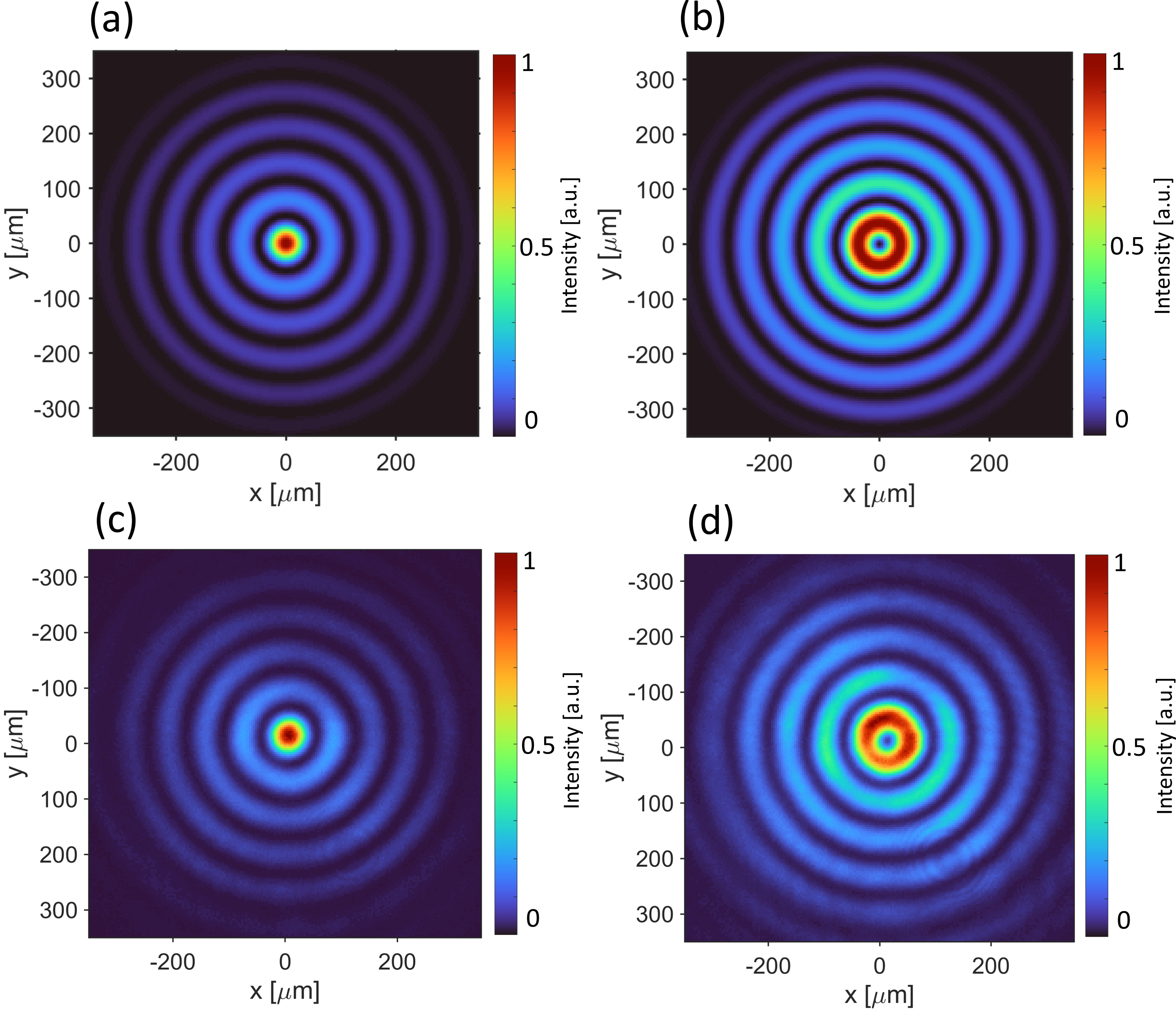}
   
   \caption{Simulated and experimentally measured transverse intensity distributions of Bessel$_0$--Gauss beams generated using the independent amplitude and phase control setup based on a single SLM, described in Fig.~\ref{fig:experimental setup}. Panels (a,b) show numerical simulations, while panels (c,d) show the corresponding experimental measurements. Panels (a,c) show the Bessel$_0$--Gauss beam, and panels (b,d) show the Bessel$_0$--Gauss beam after adding a helical phase. The helical phase produces an on-axis phase singularity, resulting in the central intensity minimum observed in panels (b,d).}
   \label{fig:bessel_gauss_helical}
\end{figure}

\begin{figure}[htp]
   \centering      
   \includegraphics[width=12.0cm]{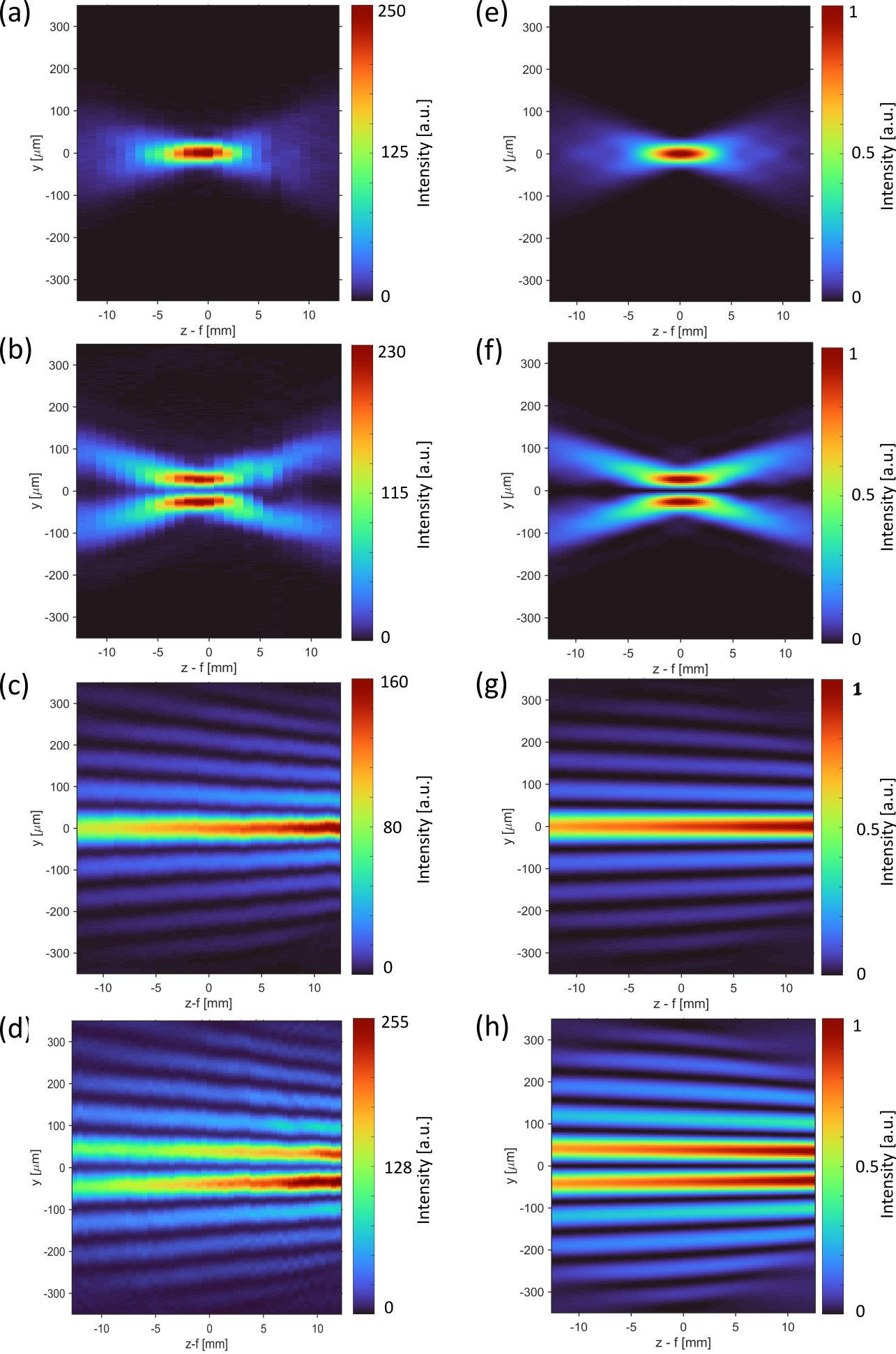}
   
   \caption{
    Experimental and simulated longitudinal intensity profiles in the $x=0$ plane.
    (a)--(d) Experimentally measured $y$--$z$ intensity distributions for, respectively, a Gaussian beam, a Laguerre--Gaussian beam with topological charge $\ell=1$, a zeroth-order Bessel--Gaussian beam, and a zero-order Bessel--Gaussian beam with a helical phase added.
    (e)--(h) Corresponding numerical simulations for the same four input fields.
    The zero-order Bessel--Gaussian beam with added helical phase provides a phase-sensitive evidence of the combined amplitude and phase modulation capability, since its generation requires both an annular amplitude profile and an azimuthally varying phase.}
    
   \label{fig:z_line}
\end{figure}

We further characterized the generated fields by recording
their longitudinal evolution around the focal region.
Figure~\ref{fig:z_line} compares measured and simulated \(y\)--\(z\) intensity
profiles for a Gaussian beam, a Laguerre--Gaussian beam with topological charge
\(\ell=1\), a Bessel$_0$--Gauss beam, and a Bessel$_0$--Gauss beam with an
added helical phase (see Appendix~\ref{app:bessel_target}). Although the
Gaussian and Bessel$_0$--Gauss beams are both centrally peaked at focus, their
axial evolution is markedly different: the Gaussian beam broadens rapidly,
whereas the Bessel$_0$--Gauss beam preserves its central structure over an
extended range. Since propagation depends on the complete complex field rather
than on the focal-plane intensity alone, the agreement between these distinct
measured and simulated axial profiles provides phase-sensitive evidence that
the programmed phase distributions are reproduced. The helical
Bessel--Gaussian case further validates simultaneous control of the nonuniform
radial amplitude and azimuthal phase.

\begin{figure}[htp]
   \centering      
   \includegraphics[width=13cm]{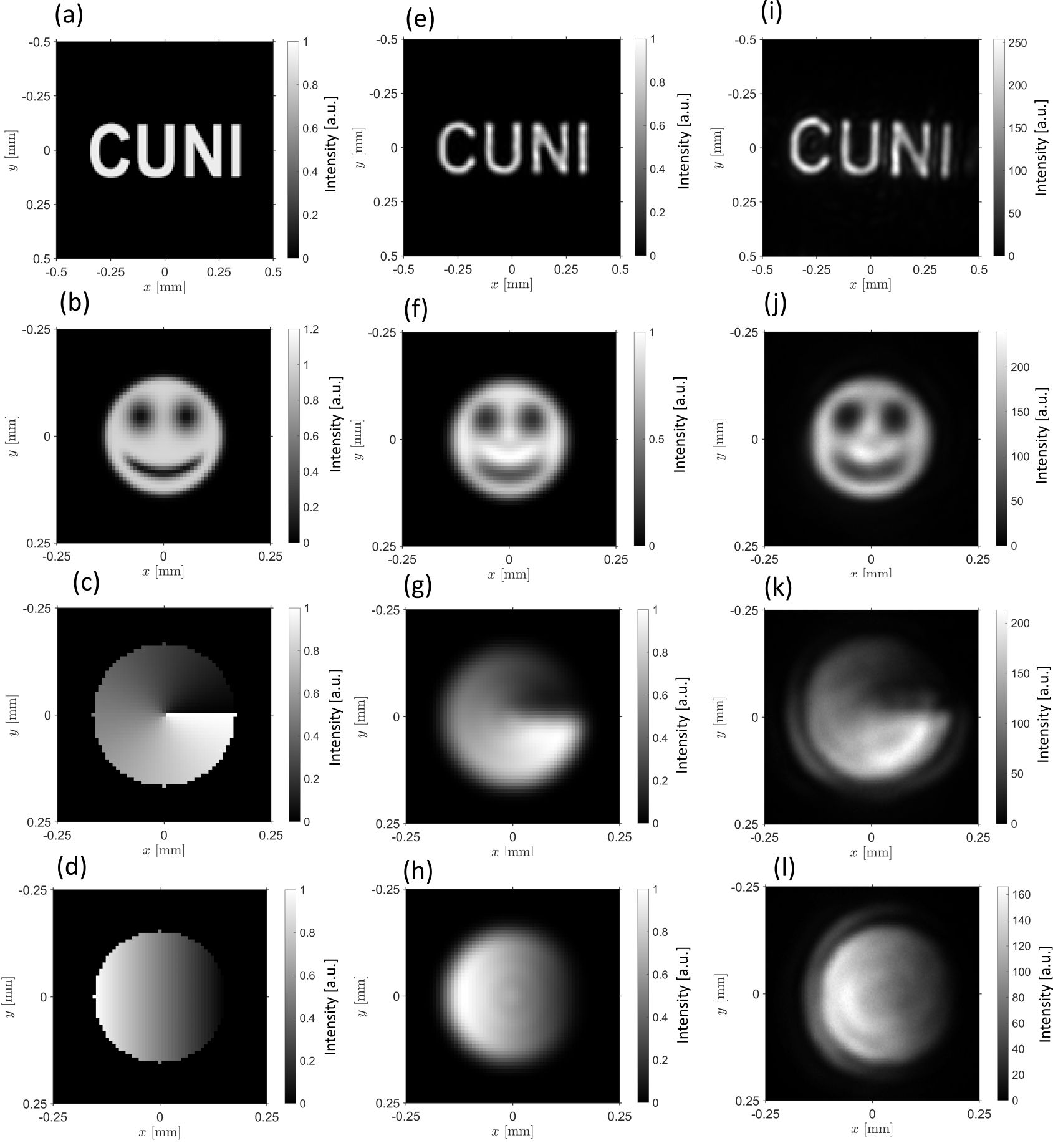}
   
   \caption{Generation of arbitrary focal-plane intensity distributions using complex-field modulation with a single SLM. Panels (a)--(d) show the target focal shapes, panels (e)--(h) show the corresponding numerical simulations, and panels (i)--(l) show the experimentally measured focal shapes. The SLM amplitude and phase masks were calculated by back-propagating the desired focal-plane fields to the pupil plane, using a spatially uniform target phase. The simulations were obtained by constructing the encoded pupil field from Eq.~\eqref{eq:two_slm_field} and propagating it to the focal plane. The helical-ramp and Zernike-tilt-like intensity profiles in panels (k) and (l) are examples of optical intensity landscapes relevant for electron-beam shaping, where the ponderomotive phase shift imprinted on the electron beam is proportional to the local optical intensity~\cite{chirita2022transverse}.}

   \label{fig:arbitrary_focal_shapes}
\end{figure}

Finally, we applied the same complex-field encoding procedure to arbitrary focal-plane intensity distributions. For these targets, the desired focal-plane amplitude was defined as \(A_{\mathrm{tar}}(x_f,y_f)=\sqrt{I_{\mathrm{tar}}(x_f,y_f)}\), with a spatially uniform target phase (see Appendix~\ref{app:arbitrary_targets}) for details. Figure~\ref{fig:arbitrary_focal_shapes} shows four representative examples: the prescribed target intensities, the corresponding simulated focal-plane intensities, and the experimentally measured patterns. The measured fields reproduce the main spatial features of the targets, including both sharp image-like structures and smooth grayscale variations. 

To quantify the agreement between the target and measured focal-plane intensity distributions in Fig.~\ref{fig:arbitrary_focal_shapes}, we used the normalized correlation coefficient~\cite{tsai2003evaluation}
\begin{equation}
    F_{\mathrm{corr}}
    =
    \frac{
    \sum_{m,n}
    \left[
    I_{\mathrm{tar}}(x_{f,m},y_{f,n})
    -
    \overline{I}_{\mathrm{tar}}
    \right]
    \left[
    I_{\mathrm{meas}}(x_{f,m},y_{f,n})
    -
    \overline{I}_{\mathrm{meas}}
    \right]
    }
    {
    \sqrt{
    \sum_{m,n}
    \left[
    I_{\mathrm{tar}}(x_{f,m},y_{f,n})
    -
    \overline{I}_{\mathrm{tar}}
    \right]^2
    \sum_{m,n}
    \left[
    I_{\mathrm{meas}}(x_{f,m},y_{f,n})
    -
    \overline{I}_{\mathrm{meas}}
    \right]^2
    }
    }.
    \label{eq:intensity_correlation}
\end{equation}
Here \(I_{\mathrm{tar}}(x_{f,m},y_{f,n})\) and \(I_{\mathrm{meas}}(x_{f,m},y_{f,n})\) are the target and measured intensities at the focal-plane pixel \((x_{f,m},y_{f,n})\), respectively, while \(\overline{I}_{\mathrm{tar}}\) and \(\overline{I}_{\mathrm{meas}}\) denote their mean values over the evaluated region. A value of \(F_{\mathrm{corr}}=1\) corresponds to perfect pixel-wise agreement between the two normalized intensity distributions, whereas smaller values indicate increasing spatial deviations.

The resulting correlation coefficients are listed in Table~\ref{tab:focal_shape_correlations}. All four measured focal-plane patterns show high agreement with their corresponding targets, with \(F_{\mathrm{corr}}\geq 0.86\). The lowest value is obtained for the letter pattern, where the sharp edges and fine spatial features are most sensitive to diffraction, alignment, and finite-resolution effects. The smoother intensity distributions yield higher correlations, reaching \(F_{\mathrm{corr}}=0.96\) for the smiley pattern.

\begin{table}[htp]
\centering
\caption{Normalized correlation coefficients between the target and experimentally measured focal-plane intensity distributions shown in Fig.~\ref{fig:arbitrary_focal_shapes}.}
\label{tab:focal_shape_correlations}
\begin{tabular}{c c c}
\hline
Target panel & Measured panel & \(F_{\mathrm{corr}}\) \\
\hline
(a) & (i) & 0.86 \\
(b) & (j) & 0.96 \\
(c) & (k) & 0.93 \\
(d) & (l) & 0.94 \\
\hline
\end{tabular}
\end{table}

\section{\label{sec:level4}Discussion and Conclusion}

We have demonstrated independent amplitude and phase control of an optical field using a single phase-only SLM. The method uses two spatially separated regions of the same device as sequential phase masks: the first region, together with a polarizer placed in the beam path, encodes the desired amplitude, while the second region imposes the target phase. Since the field produced by the first region is imaged onto the second region with unit magnification, the two masks are defined on the same pupil-plane coordinates. This provides a direct implementation of complex-field synthesis with a single reflective SLM, avoiding the need for two independent modulators~\cite{zhu2014arbitrary} while retaining the physical separation of amplitude and phase control used in two-plane approaches.

The experimental results confirm the expected behavior of the scheme. The measured intensity modulation follows the predicted \(\cos^2(\phi_{\mathrm{A}}/2)\) dependence, verifying that the first SLM region and the polarizer act as a controllable amplitude modulator. The generation of Bessel--Gaussian fields with and without an added helical phase demonstrates that the setup can independently prescribe a radial amplitude distribution and an azimuthally varying phase. The measured transverse (Fig.~\ref{fig:bessel_gauss_helical}) and longitudinal intensity (Fig.~\ref{fig:z_line}) distributions agree well with the corresponding scalar diffraction simulations, showing that the encoded complex pupil field is preserved not only in the focal plane but also through propagation around the focus.

The same encoding procedure was also applied to arbitrary focal-plane intensity patterns. In this case, the target focal fields were back-propagated to the pupil plane, normalized to the accessible amplitude range, and encoded into the two SLM regions using the direct double-phase prescription. The experimentally reconstructed patterns from Fig.~\ref{fig:arbitrary_focal_shapes} reproduce both image-like targets with sharp spatial features and smoother grayscale distributions, with normalized correlation coefficients between \(0.86\) and \(0.96\).

More generally, the method involves practical trade-offs. Direct amplitude control is intrinsically less efficient than purely phase-only beam shaping, because the polarization projection used for amplitude modulation attenuates part of the incident field. In addition, the finite pixel pitch, fill factor, phase calibration, and residual wavefront errors of the SLM limit the fidelity of high-spatial-frequency target patterns. These are implementation-specific limitations rather than fundamental restrictions of the encoding principle. The arbitrary intensity targets were encoded by back-propagating
focal-plane fields with uniform phase, chosen here for simplicity.
Because the focal-plane phase was not optimized, the resulting pupil
fields may not be optimally matched to the finite aperture, spatial
bandwidth, and incident-beam profile of the system. As a possible
extension, optimizing the target phase could therefore improve the
reconstruction fidelity and potentially reduce amplitude clipping and
the associated optical losses. Such optimization could be performed
using optimal-transport-based initial phase estimates
~\cite{swan2025high}, iterative Fourier-transform algorithms
~\cite{wyrowski1988iterative,pasienski2008high}, or cost-function-based
target optimization~\cite{bowman2017high,chirita2025design}, but is
beyond the scope of the present work.

In conclusion, we have realized a compact single-SLM platform for independent amplitude and phase shaping of optical fields. Because it provides direct complex-field control with a single programmable element, this method offers a practical route for structured illumination~\cite{hohenester2025optimizing}, holographic beam shaping~\cite{forbes2016creation,bolduc2013exact}, and optical fields designed for free-electron control~\cite{garcia2025roadmap}, including light-based electron lenses~\cite{chirita2022transverse}, aberration correction~\cite{chirita2025light}, and high purity electron vortex beams.

\appendix

\section{\label{sec:A}Jones-matrix derivation of the complex-field modulation}

The complex-amplitude modulation implemented in the present setup follows the
two-step phase-only modulation approach described in
Ref.~\cite{zhu2014arbitrary}. In this appendix, we derive the resulting field
using Jones calculus, with the single-pass optical-element matrices defined
following Ref.~\cite{hecht2023optik}. The optical arrangement is shown
schematically in Fig.~\ref{fig:experimental setup}. Throughout the derivation,
the spatial dependence of the phase patterns is retained explicitly when
needed.

Scalar propagation phases introduced by the imaging relay are
treated separately, since they do not modify the polarization state. For the
retracing section containing the quarter-wave plate and concave mirror, Jones
vectors are expressed in local right-handed transverse bases attached to the
propagation direction. The \(x\)-axis is chosen parallel to the active axis of
the SLM; upon reflection, the \(x\)-axis is retained while the orthogonal
transverse axis changes sign to preserve a right-handed coordinate system.

The incident field is assumed to be linearly polarized along the diagonal
direction, so that its Jones vector can be written as
\begin{equation}
  \mathbf{E}_{\mathrm{in}}(x_p,y_p)
  =
  \frac{E_0(x_p,y_p)}{\sqrt{2}}
  \begin{pmatrix}
    1 \\
    1
  \end{pmatrix},
  \label{eq:appendix_input_field}
\end{equation}
where \(E_0(x_p,y_p)\) denotes the complex amplitude of the incident beam
before modulation. The first addressed region of the SLM introduces a
spatially dependent phase delay \(\phi_{\mathrm{A}}(x_p,y_p)\) on the
\(x\)-polarized component, while leaving the orthogonal component unchanged.
Its Jones matrix is therefore
\begin{equation}
  \mathbf{J}_{\mathrm{SLM},1}
  =
  \begin{pmatrix}
    e^{i\phi_{\mathrm{A}}(x_p,y_p)} & 0 \\
    0 & 1
  \end{pmatrix}.
  \label{eq:appendix_slm1_matrix}
\end{equation}

After the first SLM interaction, the field becomes
\begin{equation}
  \mathbf{E}_1(x_p,y_p)
  =
  \mathbf{J}_{\mathrm{SLM},1}\mathbf{E}_{\mathrm{in}}(x_p,y_p)
  =
  \frac{E_0(x_p,y_p)}{\sqrt{2}}
  \begin{pmatrix}
    e^{i\phi_{\mathrm{A}}(x_p,y_p)} \\
    1
  \end{pmatrix}.
  \label{eq:appendix_after_slm1}
\end{equation}

The field is then projected onto the same diagonal polarization state as the
incident beam. The corresponding Jones matrix is
\begin{equation}
  \mathbf{J}_{45^\circ}
  =
  \frac{1}{2}
  \begin{pmatrix}
    1 & 1 \\
    1 & 1
  \end{pmatrix}.
  \label{eq:appendix_diagonal_polarizer}
\end{equation}

Applying this projection gives
\begin{equation}
  \mathbf{E}_2(x_p,y_p)
  =
  \mathbf{J}_{45^\circ}\mathbf{E}_1(x_p,y_p)
  =
  \frac{E_0(x_p,y_p)}{2\sqrt{2}}
  \left[
    1+e^{i\phi_{\mathrm{A}}(x_p,y_p)}
  \right]
  \begin{pmatrix}
    1 \\
    1
  \end{pmatrix}.
  \label{eq:appendix_after_polarizer}
\end{equation}

The beam subsequently passes through a quarter-wave plate,
reflects from the concave mirror, and passes through the same quarter-wave
plate in the reverse direction. After the first passage through the
quarter-wave plate, the field incident on the mirror is generally elliptically
polarized. The mirror and the reversal of the propagation direction must
therefore be included explicitly in the round-trip polarization
transformation.

For a single forward passage, the Jones matrix of the quarter-wave plate,
whose fast axis is oriented at \(\theta=\pi/8\) with respect to the \(x\)-axis,
is
\begin{equation}
  \mathbf{J}_{\lambda/4}(\theta)
  =
  \begin{pmatrix}
    \cos^2\theta+i\sin^2\theta
    &
    (1-i)\sin\theta\cos\theta
    \\
    (1-i)\sin\theta\cos\theta
    &
    \sin^2\theta+i\cos^2\theta
  \end{pmatrix}.
  \label{eq:appendix_qwp_matrix}
\end{equation}

Because the second transverse basis vector reverses after
reflection, the physical fast axis of the same quarter-wave plate is observed
at the angle \(-\theta\) in the return beam basis. The reverse-pass matrix is
therefore
\begin{equation}
\mathbf{J}_{\lambda/4}^{\mathrm{rev}}(\theta)
  =
  \mathbf{J}_{\lambda/4}(-\theta).
  \label{eq:appendix_qwp_reverse}
\end{equation}

At a finite angle of incidence, the mirror may introduce
different complex reflection factors for the two local polarization
components. In the present paraxial, near-normal-incidence geometry, this
differential diattenuation and retardance are neglected. In the beam-following
bases defined above, the mirror is consequently represented by
\begin{equation}
\mathbf{J}_{\mathrm{M}}
  \simeq
  r_{\mathrm{M}}
  \begin{pmatrix}
    1 & 0 \\
    0 & -1
  \end{pmatrix},
  \label{eq:appendix_mirror_matrix}
\end{equation}
where \(r_{\mathrm{M}}\) is the common complex reflection
coefficient. The relative minus sign accounts for the reversal of the
transverse coordinate basis, and therefore also for the reversal of
polarization handedness upon reflection; it is not an additional differential
phase introduced by the mirror coating.

The complete round-trip transformation is thus
\begin{equation}
  \begin{aligned}
    \mathbf{J}_{\mathrm{rt}}(\theta)
    &=
    \mathbf{J}_{\lambda/4}(-\theta)
    \mathbf{J}_{\mathrm{M}}
    \mathbf{J}_{\lambda/4}(\theta)
    \\
    &\simeq
    r_{\mathrm{M}}
    \begin{pmatrix}
      \cos(2\theta) & \sin(2\theta) \\
      -\sin(2\theta) & \cos(2\theta)
    \end{pmatrix}.
  \end{aligned}
  \label{eq:appendix_round_trip_matrix}
\end{equation}

Apart from the common reflection factor, the round trip
therefore acts as a rotation of the linear-polarization direction. For
\(\theta=\pi/8\),
\begin{equation}
\mathbf{J}_{\mathrm{rt}}\left(\frac{\pi}{8}\right)
  \simeq
  \frac{r_{\mathrm{M}}}{\sqrt{2}}
  \begin{pmatrix}
    1 & 1 \\
    -1 & 1
  \end{pmatrix}.
  \label{eq:appendix_hwp_matrix}
\end{equation}

After omitting the common factor \(r_{\mathrm{M}}\), this
transformation maps the diagonal polarization state onto the \(x\)-polarized
working axis of region B:
\begin{equation}
  \mathbf{E}_3(x_p,y_p)
  =
  \frac{1}{\sqrt{2}}
  \begin{pmatrix}
    1 & 1 \\
    -1 & 1
  \end{pmatrix}
  \mathbf{E}_2(x_p,y_p)
  =
  \frac{E_0(x_p,y_p)}{2}
  \left[
    1+e^{i\phi_{\mathrm{A}}(x_p,y_p)}
  \right]
  \begin{pmatrix}
    1 \\
    0
  \end{pmatrix}.
  \label{eq:appendix_after_waveplate}
\end{equation}

Residual differential reflection between the two
polarization components, an error in the quarter-wave-plate retardance, or an
angular misalignment of the wave plate would produce a small departure from
perfect \(x\)-linear polarization. These nonidealities are neglected in the
idealized Jones-matrix model.

The unit-magnification relay additionally introduces the
polarization-independent quadratic phase
\(\phi_{\mathrm{rel}}(x_p,y_p)\) defined in
Eq.~\eqref{eq:relay_phase}. Since this contribution is scalar, it does not
modify the polarization transformation described above. The coordinate
inversion produced by the relay is accounted for when constructing the phase
pattern on region B and is not written explicitly here. The field incident on
region B is therefore
\(e^{i\phi_{\mathrm{rel}}(x_p,y_p)}
\mathbf{E}_3(x_p,y_p)\).

The phase physically displayed on region B is
\(\phi_{\mathrm{B,disp}}(x_p,y_p)\), as defined in
Eq.~\eqref{eq:phase_B_displayed}. The corresponding Jones matrix is
\begin{equation}
\mathbf{J}_{\mathrm{SLM},2}
  =
  \begin{pmatrix}
    e^{i\phi_{\mathrm{B,disp}}(x_p,y_p)} & 0 \\
    0 & 1
  \end{pmatrix}
  \label{eq:appendix_slm2_matrix}
\end{equation}

The output field is consequently
\begin{equation}
  \begin{aligned}
  \mathbf{E}_{\mathrm{out}}(x_p,y_p)
  &=
  \mathbf{J}_{\mathrm{SLM},2}
  e^{i\phi_{\mathrm{rel}}(x_p,y_p)}
  \mathbf{E}_3(x_p,y_p)
  \\
  &=
  \frac{E_0(x_p,y_p)}{2}
  \left[
    1+e^{i\phi_{\mathrm{A}}(x_p,y_p)}
  \right]
  \exp\!\left\{
    i\left[
      \phi_{\mathrm{rel}}(x_p,y_p)
      +
      \phi_{\mathrm{B,disp}}(x_p,y_p)
    \right]
  \right\}
  \begin{pmatrix}
    1 \\
    0
  \end{pmatrix}
  \\
  &=
  \frac{E_0(x_p,y_p)}{2}
  \left[
    1+e^{i\phi_{\mathrm{A}}(x_p,y_p)}
  \right]
  e^{i\phi_{\mathrm{B}}(x_p,y_p)}
  \begin{pmatrix}
    1 \\
    0
  \end{pmatrix},
  \end{aligned}
  \label{eq:appendix_output_vector}
\end{equation}
where the final equality follows from
\(\phi_{\mathrm{B,disp}}
=
\phi_{\mathrm{B}}-\phi_{\mathrm{rel}}\).
Thus, the relay-induced quadratic phase is cancelled by the compensating
quadratic phase displayed on region B.

Since the output field is linearly polarized along \(x\), its scalar complex
amplitude can be written as
\begin{equation}
  E_{\mathrm{out}}(x_p,y_p)
  =
  E_0(x_p,y_p)
  \frac{
    1+e^{i\phi_{\mathrm{A}}(x_p,y_p)}
  }{2}
  e^{i\phi_{\mathrm{B}}(x_p,y_p)}.
  \label{eq:appendix_scalar_output_initial}
\end{equation}

Using the identity
\begin{equation}
  1+e^{i\phi_{\mathrm{A}}(x_p,y_p)}
  =
  2\cos\left(\frac{\phi_{\mathrm{A}}(x_p,y_p)}{2}\right)
  e^{i\phi_{\mathrm{A}}(x_p,y_p)/2},
  \label{eq:appendix_half_angle_identity}
\end{equation}
Eq.~\eqref{eq:appendix_scalar_output_initial} reduces to
\begin{equation}
  E_{\mathrm{out}}(x_p,y_p)
  =
  E_0(x_p,y_p)
  \cos\left[
    \frac{\phi_{\mathrm{A}}(x_p,y_p)}{2}
  \right]
  \exp\left\{
    i\left[
      \phi_{\mathrm{B}}(x_p,y_p)
      +
      \frac{\phi_{\mathrm{A}}(x_p,y_p)}{2}
    \right]
  \right\}.
  \label{eq:appendix_amplitude_phase}
\end{equation}

\section{\label{sec:B}Intensity modulation }
\label{Intensity modulation}
\subsection{Calibration of the amplitude modulation}\label{Intensity modulation calibration}

\begin{figure}[htp]
   \centering      
   \includegraphics[width=12cm]{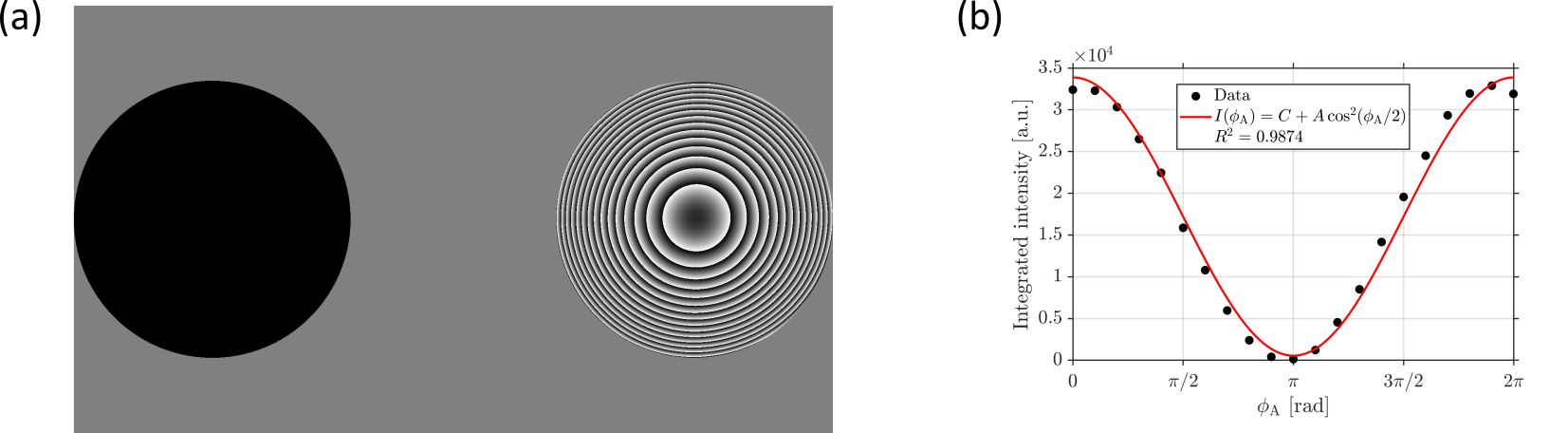}
   
   \caption{Calibration of the amplitude modulation obtained with the single-SLM complex-field encoding scheme. (a) Representative hologram applied to the SLM during the calibration. The left SLM region is used for amplitude modulation by applying a spatially uniform phase \(\phi_{\mathrm{A}}\), while the right region contains the phase pattern \(\phi_{\mathrm{B}}\), here chosen as a lens function. (b) Measured integrated intensity on the CMOS detector as a function of the phase shift applied to the first modulation region. }
\label{fig:hologram_intensity_calibration}
\end{figure}

To verify the amplitude response of the two-region SLM encoding scheme, we performed a calibration measurement using the setup shown in Fig.~\ref{fig:experimental setup}. Region B was programmed with a quadratic lens phase,
\begin{equation}
    \phi_{\mathrm{lens}}(x_p,y_p)
    =
    -\frac{\pi}{\lambda_\mathrm{L} f_m}
    \left(x_p^2+y_p^2\right),
    \label{eq:lens_phase}
\end{equation}
where \(f_m=250~\mathrm{mm}\). This phase profile was used to collimate the beam after reflection from the SLM. Region A was addressed with a spatially uniform phase offset, $\phi_{\mathrm{A}}$, which was varied from 0 to approximately $2\pi$. The resulting beam intensity was recorded on a CMOS camera.
 A representative hologram is shown in Fig.~\ref{fig:hologram_intensity_calibration}(a), and the corresponding integrated Gaussian intensity is plotted in Fig.~\ref{fig:hologram_intensity_calibration}(b). As expected from Eq.~\eqref{eq:two_slm_field}, the field amplitude scales as $\cos(\phi_{\mathrm{A}}/2)$, giving an intensity dependence $I(\phi_{\mathrm{A}})=C+A\cos^2(\phi_{\mathrm{A}}/2)$. The fit yields $R^2=0.9874$, indicating that the $\cos^2(\phi_{\mathrm{A}}/2)$ model accounts for 98.74\% of the measured intensity variance. This agreement verifies the expected amplitude-modulation response, while the remaining deviations are attributed to residual alignment errors, imperfect polarizer and wave-plate angles, finite extinction ratio, and SLM phase-calibration errors.

\subsection{Image-plane resolution of the amplitude modulation}
\label{sec:spatial_resolution}

\begin{figure}[htp]
   \centering      
   \includegraphics[width=12cm]{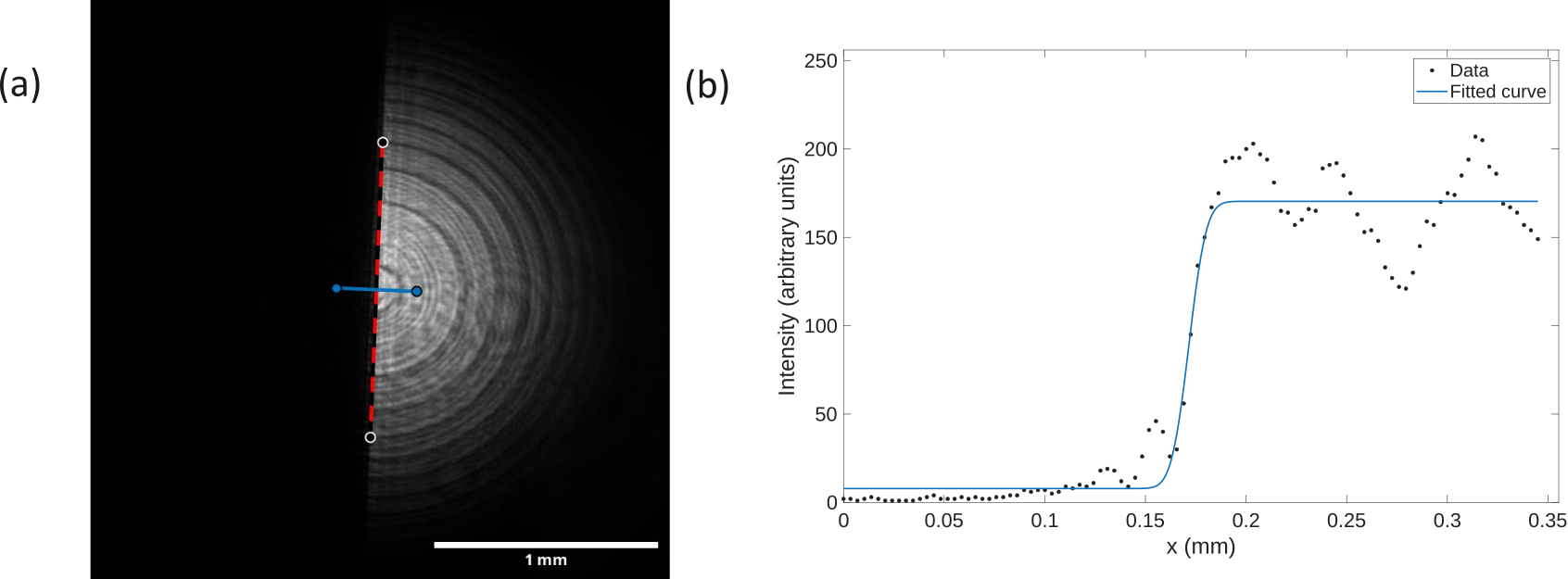}
   
   \caption{Resolution measurement of the amplitude-modulation step. (a) Measured intensity in the image plane for a split transmission pattern applied to region \(A\) of the SLM, with zero transmission on one half of the region and maximum transmission on the other half. The red dashed line indicates the edge position selected manually, while the blue line marks the perpendicular direction along which the intensity profile was sampled. (b) Intensity profile extracted along the blue line in (a). The data were fitted with an error function, from which the \(10\%\)--\(90\%\) rise distance of the imaged amplitude edge was determined.}
   \label{fig:resolution_measurement}
\end{figure}

The spatial resolution of the amplitude-modulation step was evaluated by imaging a sharp amplitude edge produced on region A of the SLM. For this measurement, the CMOS camera was moved from the Fourier plane to the image plane of the SLM, so that the modulated intensity distribution could be recorded directly. A split pattern was applied to region A, with zero transmission on one half of the region and maximum transmission on the other half. The input beam diameter was reduced by a factor of two compared with the other measurements in this manuscript in order to minimize optical artefacts. The resulting image-plane intensity distribution is shown in Fig.~\ref{fig:resolution_measurement}(a). The edge position was selected manually, and an intensity profile was extracted perpendicular to the edge. Fitting this profile with an error function, as shown in Fig.~\ref{fig:resolution_measurement}(b), gave a \(10\%\)--\(90\%\) rise distance of \(\SI{18.4}{\micro\metre}\). The oscillations near the edge are consistent with diffraction from the sharp amplitude step. This measured resolution is specific to the present optical alignment and imaging conditions, and should not be interpreted as a fundamental limitation of the complex-field modulation method.

\subsection{General double-phase hologram formalism}
\label{app:general_hologram_formalism}

All focal-plane holograms were generated using the same numerical
procedure. A desired complex field
\begin{equation}
    E_{\mathrm{tar}}(x_f,y_f)
    =
    A_{\mathrm{tar}}(x_f,y_f)
    \exp[i\Phi_{\mathrm{tar}}(x_f,y_f)]
\end{equation}
was first specified in the focal plane of the Fourier lens. For a
square computational grid, the focal-plane sampling is
\begin{equation}
    \Delta x_f
    =
    \frac{\lambda_{\mathrm{L}} f_{\mathrm{F}}}
    {N_{\mathrm{calc}}p},
    \label{eq:focal_sampling}
\end{equation}
where $\lambda_{\mathrm{L}}$ is the optical wavelength,
$f_{\mathrm{F}}$ is the focal length of the Fourier lens, $p$ is the
SLM pixel pitch, and $N_{\mathrm{calc}}$ is the computational grid
size.

In the calculations presented here, $N_{\mathrm{calc}}=2414$, while
the addressed pupil occupies $N_{\mathrm{p}}=700$ pixels along each
transverse direction. The pupil is therefore embedded in a larger
computational grid, providing zero padding and a focal-plane sampling
of approximately \SI{8.0}{\micro\meter}. This zero padding refines the
numerical sampling of the focal-plane distribution but does not change
the physical spatial resolution.

The circular pupil has a pixel radius $R_{\mathrm{px}}=350$ and a
physical radius
\begin{equation}
    a=R_{\mathrm{px}}p=\SI{2.8}{mm},
\end{equation}
corresponding to a pupil diameter $D=2a=\SI{5.6}{mm}$. The finite pupil
sets a characteristic focal-plane feature size of order
\begin{equation}
    \delta x_f
    \sim
    \frac{\lambda_{\mathrm{L}}f_{\mathrm{F}}}{D},
    \label{eq:focal_resolution}
\end{equation}
which is approximately \SI{28}{\micro\meter} for
$\lambda_{\mathrm{L}}=\SI{1030}{nm}$ and
$f_{\mathrm{F}}=\SI{150}{mm}$. The SLM pixel pitch $p$ sets the maximum transverse spatial frequency
that can be represented without aliasing. Through the Fourier-lens
mapping, this corresponds to an alias-free focal-plane half-width
\begin{equation}
    x_{f,\max}
    \simeq
    \frac{\lambda_{\mathrm{L}}f_{\mathrm{F}}}{2p}
    \approx
    \SI{9.7}{mm}.
    \label{eq:focal_fov}
\end{equation}
Consequently, sharp target features below the supported transverse
resolution are smoothed by the finite pupil. Under ideal
unit-magnification imaging, the two-region encoding introduces no
additional fundamental spatial-frequency cutoff, although relay
aberrations, residual registration errors, and the relay
modulation-transfer function may further reduce the fidelity of
high-spatial-frequency features.

The pupil-plane field required to generate the target focal field was
obtained by inverse Fourier propagation and restriction to the
illuminated circular pupil,
\begin{equation}
    P_{\mathrm{des}}(x_p,y_p)
    =
    \Pi_a(x_p,y_p)
    \mathcal{F}^{-1}
    \left\{
        E_{\mathrm{tar}}(x_f,y_f)
    \right\},
    \label{eq:desired_pupil_field}
\end{equation}
where $\Pi_a(x_p,y_p)$ is the circular pupil function of physical
radius $a$. Writing
\begin{equation}
    P_{\mathrm{des}}(x_p,y_p)
    =
    A_{\mathrm{des}}(x_p,y_p)
    \exp[i\Phi(x_p,y_p)],
    \label{eq:desired_pupil_amp_phase}
\end{equation}
the amplitude available before the double-phase modulation is
\begin{equation}
    A_{\mathrm{avail}}(x_p,y_p)
    =
    \Pi_a(x_p,y_p)A_{\mathrm{in}}(x_p,y_p),
    \label{eq:available_amplitude}
\end{equation}
where $A_{\mathrm{in}}(x_p,y_p)$ is the incident beam amplitude.

The normalized amplitude encoded by region A was chosen as
\begin{equation}
    A_{\mathrm{enc}}(x_p,y_p)
    =
    \eta
    \frac{A_{\mathrm{des}}(x_p,y_p)}
         {A_{\mathrm{avail}}(x_p,y_p)},
    \qquad
    0\leq A_{\mathrm{enc}}\leq 1,
    \label{eq:encoded_amplitude_general}
\end{equation}
within the illuminated pupil. Here, $\eta$ is the largest global
scaling factor, including a small numerical safety margin, for which
the encoded amplitude remains within the allowed range. This
normalization accounts for both the incident Gaussian-beam profile and
the finite pupil transmission.

The complex pupil field is encoded using the two phase-only SLM
regions. Before inclusion of the relay and wavefront corrections, the
field after the two modulation steps is
\begin{equation}
    E_{\mathrm{out}}(x_p,y_p)
    =
    A_{\mathrm{avail}}(x_p,y_p)
    \cos\left[
        \frac{\phi_{\mathrm{A}}(x_p,y_p)}{2}
    \right]
    \exp\left\{
        i\left[
            \frac{\phi_{\mathrm{A}}(x_p,y_p)}{2}
            +
            \phi_{\mathrm{B}}(x_p,y_p)
        \right]
    \right\}.
    \label{eq:double_phase_general}
\end{equation}
The effective phase masks are chosen as
\begin{equation}
    \phi_{\mathrm{A}}(x_p,y_p)
    =
    2\cos^{-1}
    \left[
        A_{\mathrm{enc}}(x_p,y_p)
    \right],
    \label{eq:general_phi_A}
\end{equation}
and
\begin{equation}
    \phi_{\mathrm{B}}(x_p,y_p)
    =
    \Phi(x_p,y_p)
    -
    \frac{\phi_{\mathrm{A}}(x_p,y_p)}{2}.
    \label{eq:general_phi_B}
\end{equation}
Substitution into Eq.~\eqref{eq:double_phase_general} gives
\begin{equation}
    E_{\mathrm{out}}(x_p,y_p)
    =
    \eta
    P_{\mathrm{des}}(x_p,y_p),
\end{equation}
so that the desired pupil field is reproduced up to the global
amplitude factor $\eta$.

The phase physically displayed on region B additionally includes the
compensation of the quadratic phase introduced by the imaging relay,
\begin{equation}
    \phi_{\mathrm{B,disp}}(x_p,y_p)
    =
    \phi_{\mathrm{B}}(x_p,y_p)
    -
    \phi_{\mathrm{rel}}(x_p,y_p),
    \label{eq:general_displayed_phase_B}
\end{equation}
where $\phi_{\mathrm{rel}}$ is defined in
Eq.~\eqref{eq:relay_phase}. After
propagation through the relay, the contributions
$\phi_{\mathrm{rel}}$ and $-\phi_{\mathrm{rel}}$ cancel, leaving the
effective phase $\phi_{\mathrm{B}}$ used in
Eq.~\eqref{eq:double_phase_general}.

Both phase masks were wrapped modulo $2\pi$. The resulting absolute
phase masks were converted into additive phase corrections relative to
the uniform SLM background phase and cropped to the central
$N_{\mathrm{p}}\times N_{\mathrm{p}}$ region before being inserted into
regions A and B of the SLM. Additional residual aberrations were
compensated by superimposing Zernike phase-correction patterns
~\cite{lakshminarayanan2011zernike} on region B.

\subsection{Focal-plane target fields}
\label{app:focal_target_fields}

The different holograms used in this work differ only in the definition of the target focal-plane field $E_{\rm tar}$. The subsequent inverse propagation and double-phase encoding were identical for all targets.

\subsubsection{Bessel target}
\label{app:bessel_target}

For the Bessel beam, the target field was prescribed analytically in the focal plane as
\begin{equation}
    E_{\rm tar}(\rho_f)
    =
    C
    J_0(k_r\rho_f)
    \exp\left[
    -\left(\frac{\rho_f}{w}\right)^8
    \right]
    W(\rho_f),
    \label{eq:bessel_target_compact}
\end{equation}
where \(\rho_f=\sqrt{x_f^2+y_f^2}\) is the radial coordinate in the focal plane, \(J_0\) is the zeroth-order Bessel function of the first kind that defines the underlying Bessel-beam radial profile~\cite{durnin1987exact}, \(C\) is a normalization constant, \(w=350~\mu{\rm m}\) is the super-Gaussian envelope radius, and \(W(\rho_f)\) is a finite numerical window with a smooth edge taper. The transverse wave number was chosen such that the first zero of \(J_0\) occurred at
\begin{equation}
    \rho_{f,1} = 50~\mu{\rm m}.
\end{equation}
Thus,
\begin{equation}
    k_r
    =
    \frac{j_{0,1}}{\rho_{f,1}}
    =
    \frac{2.4048}{50~\mu{\rm m}}
    \simeq
    4.81\times10^4~{\rm m}^{-1},
\end{equation}
where \(j_{0,1}\) is the first zero of \(J_0\). The eighth-order super-Gaussian envelope was used to keep the central Bessel rings nearly unattenuated while smoothly suppressing the field at large radius. The implemented target is therefore a finite-aperture, super-Gaussian-apodized Bessel beam rather than an ideal infinite-energy Bessel beam.

\subsubsection{Arbitrary focal-shape targets}
\label{app:arbitrary_targets}

For arbitrary focal shapes, the target was specified through a desired focal-plane intensity distribution $I_{\rm tar}(x_f,y_f)$. This intensity could be generated analytically, for example as a ramp or Zernike-like pattern, or supplied numerically as a rendered text pattern, logo, or custom grayscale image. The corresponding complex target field was taken as
\begin{equation}
    E_{\rm tar}(x_f,y_f)
    =
    \sqrt{I_{\rm tar}(x_f,y_f)}
    \exp(i\phi_0),
    \label{eq:arbitrary_target_compact}
\end{equation}
where $\phi_0$ is a constant prescribed focal-plane phase. Thus, the target phase was taken to be spatially uniform, while the desired information was encoded in the focal-plane intensity. For the correlation analysis, the measured CMOS
intensity was normalized and linearly resampled onto the physical
coordinate grid of the target distribution, followed by translational
registration to account for camera alignment. The target distribution
was not upsampled to the camera resolution. The reported correlation
coefficient therefore quantifies the overall agreement between the
measured and prescribed intensity distributions over the same physical
field of view and is not interpreted as a measure of the optical
resolution.

For text or image-based targets, the grayscale image was first rendered on a finite focal-plane crop, normalized to unit maximum intensity, optionally flipped to match the experimental camera orientation, and thresholded to remove weak background contributions. The processed intensity was then inserted into the full computational grid and multiplied by the same finite-support focal-plane window. The resulting complex field was passed to the general inverse-propagation and double-phase encoding procedure described in Sec.~\ref{app:general_hologram_formalism}.

\begin{backmatter}
\bmsection{Funding}
M.C.C.M. acknowledges the support of the project MSCA Fellowships \newline CZ--UK4
(CZ.02.01.01/00/22\_010/0013392) under the Programme Johannes Amos Comenius.
M.K. acknowledges the support of the Czech Science Foundation (project 22-13001K),
the European Union (ERC, eWaveShaper, 101039339), and the Ministry of Education, Youth
and Sports of the Czech Republic (TERAFIT project No.\ CZ.02.01.01/00/22\_008/0004594,
funded by OP JAK, call Excellent Research).

\bmsection{Author contributions}
MCCM (lead) and MK conceived the experiment, interpreted the results and supervised the project. MCCM (lead) and MF built and optimized the experimental setup, acquired the data, and performed the experimental analysis. MCCM performed the numerical simulations and wrote the first draft of the manuscript. All authors revised the manuscript, and contributed to the final version.

\bmsection{Disclosures}
The authors declare no conflicts of interest.

\bmsection{Data availability Statement} The data supporting the findings of this study are openly available at~\cite{ChiritaMihaila2026Zenodo}.

\end{backmatter}

%%%%%%%%%% If using BibTeX:

\bibliography{literatur}

\end{document}